\documentstyle[epsfig,12pt]{article}  
\newcommand{\beq}{\begin{equation}}
\newcommand{\eeq}{\end{equation}}
\newcommand{\bea}{\vspace{0.25cm}\begin{eqnarray}}
\newcommand{\eea}{\end{eqnarray}}

\newcommand{\qb}{\mbox{{\bf
q}}}
\newcommand{\pb}{{{\bf p}}}

\newcommand{\kb}{\mbox{{\bf k}}}

\def\lsim{\mathrel{\rlap{\lower4pt\hbox{\hskip1pt$\sim$}}
    \raise1pt\hbox{$<$}}}         
\def\gsim{\mathrel{\rlap{\lower4pt\hbox{\hskip1pt$\sim$}}
    \raise1pt\hbox{$>$}}}         
\begin{document}
\vspace*{-2cm}
 
\bigskip

\begin{center}

\renewcommand{\thefootnote}{\fnsymbol{footnote}}

  {\Large\bf
Jet color chemistry and anomalous baryon production 
in $AA$-collisions}
\\
\vspace{.7cm}
\renewcommand{\thefootnote}{\arabic{footnote}}
\medskip
  {\large
  P.~Aurenche$^a$ and B.G.~Zakharov$^{b}$}
  \bigskip

{\it
$^{a}$
LAPTH, Universit\'e de Savoie, CNRS,\\
BP 110, F-74941, Annecy-le-Vieux Cedex, France\\
$^{b}$L.D. Landau Institute for Theoretical Physics,
        GSP-1, 117940,\\ Kosygina Str. 2, 117334 Moscow, Russia\\
\vspace{1.7cm}}
  {\bf
  Abstract}
\end{center}
{
\baselineskip=9pt
We study anomalous high-$p_T$ baryon production in $AA$-collisions
due to formation of the two parton collinear $gq$ system in the anti-sextet color state for quark jets 
and $gg$ system in the decuplet/anti-decuplet color states for gluon jets. 
Fragmentation of these states, which are absent for $NN$-collisions, after escaping from 
the quark-gluon plasma leads to baryon production. Our qualitative estimates
show that this mechanism can be potentially important at RHIC and LHC
energies.
\vspace{.5cm}
}

\section{Introduction}
\label{intro}
One of the intriguing results from the RHIC program (for a review, 
see \cite{MT})
is the significant enhancement of the baryon/meson ratios in $AA$-collisions
at $1.5\lsim p_{T}\lsim 5$ GeV \cite{PHENIX1,STAR1,STAR2}, 
the so-called
``baryon anomaly''.
The underlying mechanisms of this effect remain not well understood. 
It has been suggested \cite{SJL1,SJL2} that the physical reason for 
the baryon anomaly may lie in the enhancement of the string junction (SJ) loops
in the quark-gluon plasma (QGP) produced in $AA$-collisions. The notion of SJ was introduced by
Rossi and Veneziano \cite{RV1,RV2} in the generalization of the topological
expansion scheme \cite{top-exp} to the processes with baryons.
In the Rossi-Veneziano approach \cite{RV1,RV2} the baryons have a 
$Y$-shaped string configuration, and the string junction trajectory is the
intersection of three planar sheets. 
The SJ concept arises naturally from the gauge-invariant
expression of the operator with baryon quantum numbers
\bea  
B_{\alpha\beta\gamma}=
\epsilon^{i_{1}i_{2}i_{3}}
\Big[P\exp\int_{p(x,x_{1})} A_{\mu}dx^{\mu} q_{\alpha}(x_{1})\Big]_{i_{1}}
\Big[P\exp\int_{p(x,x_{2})} A_{\mu}dx^{\mu} q_{\beta}(x_{2})\Big]_{i_{2}}\nonumber\\
\times\Big[P\exp\int_{p(x,x_{3})} A_{\mu}dx^{\mu}
  q_{\gamma}(x_{3})\Big]_{i_{3}}
\,.
\label{eq:10}
\eea
The $Y$-configuration for baryons is now supported by the lattice simulation \cite{Y-shape1,Y-shape2}. 
But there have not so far been proposed dynamical computational methods 
for the SJ effects neither in high energy hadron collisions nor in 
hadronization of the expanding QGP.
For this reason the SJ model \cite{SJL1} remains highly speculative, 
and the connection of the SJ loops to the baryon anomaly needs elucidation. 

In the model of SJ loops
\cite{SJL1} the baryon/meson enhancement in $AA$-collisions 
is not a fragmentation effect, but 
is a specific soft effect of plasma hadronization extending to moderate
$p_{T}$. Another soft mechanism, which can
potentially contribute to the anomalous baryon production in $AA$-collisions,
is the quark recombination/coalescence \cite{REC_H,REC_F,REC_L} 
(for a review, see \cite{REC_rev}). 
Since baryons produced via a quark clustering receive larger 
transverse momentum than mesons it can lead to an enhancement of 
the baryon/meson ratios. The recombination models have had considerable 
success in explaining
the dependence of the hadron spectra on the number of constituent quarks.
Unfortunately, the recombination picture does not have 
a firm first-principle theoretical justification at present. 
In particular, the possibility of neglecting 
parton interactions in the hadronization process, which is the basic 
assumption of the recombination model, remains an open question.
The mechanism of the quark dominance in the clustering process
remains unclear as well.

In the present paper we discuss a mechanism for the anomalous baryon
production, which, contrary to the SJ loops and recombination
mechanisms, is related to the color dynamics of fast partons, and persists
at high $p_{T}$. It 
has been proposed in \cite{M98}, 
several years before the experimental observation of 
the baryon anomaly at RHIC.
In this mechanism the baryons are product of the hadronization
of the collinear two parton systems escaping from the QGP in the
anomalous color states:
the color anti-sextet $gq$ system for quark jets, and
the color decuplet/anti-decuplet $gg$ system for gluon jets.
To leading order in the coupling constant
these two parton states can be produced
via the processes $q\rightarrow |qg\rangle_{\{\bar{6}\}}$ and $g\rightarrow |gg\rangle_{\{10\},\{\overline{10}\}}$,
which are possible in the presence of the QGP due to color exchanges 
between the fast moving partons and the thermal partons. 
The
$|qg\rangle_{\{\bar{6}\}}$ and $|gg\rangle_{\{10\},\{\overline{10}\}}$
states can emerge from the $gq$ and $gg$ systems 
originally produced either via the vacuum (DGLAP) or induced
gluon emission \cite{BSZ}.
After escaping from the QGP 
the color anomalous two parton states result in creation of color tubes 
attached to these systems with the same anomalous color fluxes. 
The breaking of these color flux tubes 
via the Schwinger tunnel $q\bar{q}$ creation produces the hadron
systems containing baryons. For the two parton states with not very
soft gluons the baryons will be produced in the hard jet fragmentation
region.
We will call this mechanism the color anomalous baryon fragmentation (CABF).

In the CABF the production of a baryon in the jet fragmentation is 
compensated by production of an anti-baryon in the small momentum
QGP hadronization region. Thus one can say that the color exchanges with 
the QGP catalyze the baryon number flow over large rapidity gap inside jets.
Physically, it is analogous to the baryon number flow in the 
model of baryon-anti-baryon 
annihilation and baryon stopping developed in \cite{DEC,Y1988,Y1989,bflow}
(for a review, see \cite{ann_rev}).
In this model $p\bar{p}$-annihilation at the energies about
several tens of GeV is related to the diquark transition from
$\{\bar{3}\}$ to $\{6\}$ color state due to gluon exchange. This mechanism
gives natural explanation why in the experimentally studied energy 
region of $p\bar{p}$-annihilation
$\sigma_{\bar{p}p}^{ann}\propto s^{-1/2}$ \cite{Y1989}. 
The diquark transition $D_{\{3\}}\to D_{\{\bar{6}\}}$ also leads to a strong 
baryon stopping \cite{bflow}. This mechanism
explains  the slow rapidity dependence $\propto \exp(-y/2)$ of
the baryon stopping observed at ISR \cite{ISR}. This effect cannot be explained
in the standard 
quark-gluon string model \cite{MQGS1,MQGS2,LUND} which treats the diquark
as a point-like color anti-triplet object. 
In the model \cite{ann_rev} the dependence 
$\sigma_{\bar{p}p}^{ann}\propto s^{-1/2}$ is a pre-asymptotic effect, and 
not the real limiting behavior of the annihilation cross section.
In the high energy limit the baryon number flow in $\bar{p}p$-annihilation 
is related to the color decuplet double gluon exchange. It leads to 
the three sheet events with the energy independent cross section 
$\sim 1-2$ mb \cite{DEC,Y1988}.
The analysis of the experimental data on the difference $\Delta \sigma_{n}=
\sigma_{n}^{\bar{p}p}-\sigma_{n}^{pp}$ between the topological cross sections
at the energies 10-1500 GeV indeed shows that the annihilation cross section
has a small almost energy independent component
$\sigma^{\bar{p}p}_{ann}\sim 1.5$ mb \cite{DEC}.
From the point of view of the SJ concept the annihilation via 
the color decuplet $gg$-exchange
can be regarded as a pQCD analogue of the $J\bar{J}$-annihilation.   
Gotsman and Nussinov \cite {GN} have argued that 
due to the vector nature of the
gluon field $\sigma_{J\bar{J}}^{ann}$ should be
approximately energy independent, and  
from the purely geometrical estimates obtained $\sigma_{J\bar{J}}^{ann}\sim
1-2$ mb. It agrees well with the contribution
of the color decuplet double gluon mechanism obtained in \cite{DEC,Y1988}.
In pQCD the energy independence of the annihilation cross section
for the decuplet double gluon mechanism is also a direct consequence of 
the unit gluon spin.  
In the sense of the unitarity condition 
$\sigma_{J\bar{J}}^{ann}\approx \mbox{const}$ would correspond 
to the $J\bar{J}$ Regge trajectory (if it exists) with intercept 
$\alpha_{J\bar{J}}(0)\approx
1$. It contradicts to the Rossi and Veneziano hypothesis that
$\alpha_{J\bar{J}}(0)\approx 0.5$ and that namely $J\bar{J}$-annihilation 
is responsible for $\sigma_{\bar{p}p}^{ann}\propto
s^{-1/2}$ at energies $\sim 10$ GeV. However, the Rossi-Veneziano scenario 
leads to several apparent contradictions \cite{ann_rev} that 
have not satisfactory explanations within the topological expansion scheme
\cite{RV1,RV2} (the interested reader is referred to \cite{ann_rev} for details).
The problems inherent to the Rossi-Veneziano scenario 
with $\alpha_{J\bar{J}}(0)\approx 0.5$ do not arise in the scenario with the 
gluon mechanism of the baryon number flow at the pre-asymptotic
energies \cite{Y1989,ann_rev}. From the point of view of this model the
physical mechanism of the pre-asymptotic $p\bar{p}$-annihilation 
lies outside the topological expansion and has nothing 
to do with the $J\bar{J}$ Regge trajectory.
We emphasize that $p\bar{p}$-annihilation 
due to the $D_{\{3\}}\to D_{\{\bar{6}\}}$ transition
\cite{Y1989} does not contradict 
to the $Y$-shaped baryon configuration. But it shows that the corrections
to the topological expansion scheme may be huge. 
In principle, strong deviation from the topological expansion in the high energy
processes with baryons is not surprising. Indeed, in this approach the SJ is a 
point-like object, but the lattice simulation 
shows that the SJ is a highly non-local object with the
size $\sim 1$ fm \cite{Y-shape2}.

Although, the gluon mechanism of the baryon number 
flow \cite{DEC,Y1988,Y1989,bflow} was successful in explaining 
a variety of the experimental data on the processes with baryons,
the estimates of its contribution to
the baryon production in $AA$-collisions are still lacking.
A reliable quantitative computation of the CABF 
can be hardly made at present.
But as a first step toward 
understanding the potential importance of the
CABF at RHIC and LHC, it would be interesting to make at 
least rough estimates.
This is the purpose of the present paper.

The plan of the paper is as follows. 
In Sec.~2 we formulate our basic assumptions on the medium modification
of the high-$p_{T}$ hadron spectra. In Sec.~3 we discuss the physical
picture of the CABF related to the
parton processes $q\rightarrow |qg\rangle_{\{\bar{6}\}}$ and $g\rightarrow
|gg\rangle_{\{10\},\{\overline{10}\}}$,
and then give our basic formulas for the 
anomalous baryon fragmentation.
In Sec.~4  we present the numerical results. 
We give conclusions in Sec.~5.

\section{High-$p_{T}$ hadron yields and the medium-modified 
fragmentation functions}

Let us begin with our basic assumptions about the medium modification of the
high-$p_T$ hadron production.
We assume that the differential yield of 
high-$p_{T}$ hadrons in $AA$-collisions can be written in terms of  
the medium-modified cross section for the $N+N\rightarrow h+X$ process \cite{BDMS_RAA}.
In analogy to the ordinary pQCD, we write it in the form
\beq
\frac{d\sigma_{m}(N+N\rightarrow h+X)}{d\pb_{T} dy}=
\sum_{i}\int_{0}^{1} \frac{dz}{z^{2}}
D_{h/i}^{m}(z, Q,L)
\frac{d\sigma(N+N\rightarrow i+X)}{d\pb_{T}^{i} dy}\,.
\label{eq:20}
\eeq
Here $\pb_{T}^{i}=\pb_{T}/z$ is the parton transverse momentum,
${d\sigma(N+N\rightarrow i+X)}/{d\pb_{T}^{i} dy}$ is the ordinary
hard cross section, and $D_{h/i}^{m}$ is now the medium-modified fragmentation 
function (FF) for transition of an initial parton $i$ to 
the observed hadron $h$, $L$ is the parton path length in the QGP. 

As usual we assume that in $AA$-collisions hadronization
of fast partons happens after escaping from the QGP.
Under the usual assumption of 
independent parton fragmentation/hadronization in vacuum
the $D_{h/i}^{m}$ can be approximated as \cite{raa08}
\beq
D_{h/i}^{m}(z,Q,L)\approx \int_{z}^{1} \frac{dz'}{z'}D_{h/j}(z/z',Q(L))
D_{j/i}^{m}(z',Q,L)\,,
\label{eq:30}
\eeq
where  there is an implicit sum over the parton species $j=q,g$,  
$D_{h/j}(z,Q(L))$ is the ordinary vacuum FF, 
and
$D_{j/i}^{m}(z',Q,L)$ is the medium-modified FF 
for transition of an initial parton $i$ with virtuality $Q$
to a parton $j$ escaping from the QGP. Here $Q(L)$
is the typical virtuality of the final partons at the boundary of the QGP. 
From the uncertainty relation $\Delta E\Delta t\gsim 1$
one can obtain 
$Q^{2}(L)\sim \max{(Q/L,Q_0^{2})}$, where 
$Q_{0}\sim 1-2$ GeV is some minimal nonperturbative scale.
We will discuss jets with energy up to several tens of GeV.
For RHIC and LHC conditions the lifetime/size of the QGP
is about the nucleus radius, $R_A$.
For $L\sim R_{A}\sim 6$ fm from the above formula one can see that 
for parton energy $E\lsim 100$ GeV
the
fragmentation/hadronization scale turns out to be relatively small 
$\mu_{h}\sim Q(L)\sim Q_{0}$. For such a virtuality scale the parton
fragmentation outside the QGP can be described as hadronization of the
color flux tubes/strings attached to the fast partons
\cite{Nussinov,MQGS1,MQGS2,LUND},
which comes through sequential breaking of 
the color flux tubes via the Schwinger tunnel  production of $q\bar{q}$ pairs.

The medium-modified FF $D_{j/i}^{m}(z,Q,L)$ accumulates both the ordinary
vacuum (DGLAP) gluon emission and the induced gluon radiation due to multiple
scattering in the QGP. 
In general, one cannot separate the DGLAP and the induced stages since at
least part of the DGLAP shower occurs in the medium.
However, in the parton energy region $E\lsim 100$ GeV the effect of the
overlapping of these two stages should
not be strong. The point is that at such energies almost all the DGLAP gluon emission 
occurs at the length/time scale 
$\lsim 0.3-1$ fm \cite{raa08}. It is of the order of the QGP formation 
time $\tau_{0}\sim 0.5$ fm. For this reason in a first approximation one can neglect the overlapping of
the DGLAP region with the medium. 
Then taking advantage of the fact that the induced gluon emission occurs
mostly after the DGLAP stage, in the length range 
approximately from $\tau_{0}$ to $R_{A}$, 
one can write approximately the medium-modified FF $D_{j/i}^{m}$ 
as a convolution \cite{raa08}
\beq
D_{j/i}^{m}(z,Q,L)\approx\int_{z}^{1} \frac{dz'}{z'}D_{j/l}^{ind}(z/z',E_{l},L)
D_{l/i}^{DGLAP}(z',Q)\,,
\label{eq:31}
\eeq
where $E_{l}=Qz'$,
$D_{j/l}^{ind}$ is the induced FF for parton transition $l\to j$ in the QGP
(it depends on the parton energy $E$, but not the virtuality), and 
$D_{l/i}^{DGLAP}$ is the ordinary DGLAP partonic FF.

In the above physical picture with independent parton fragmentation 
the major medium 
effect is a shift of the variable $z$ in $D_{j/i}^{m}(z,Q,L)$  in (\ref{eq:31})
as compared to the variable $z'$ in the DGLAP FF in (\ref{eq:31})
due to the induced parton energy loss described by the  
$D_{j/l}^{ind}$ FF. This leads to suppression of the hadron spectra
in $AA$-collisions (usually called the jet quenching). 
Because the hard cross sections
are steeply falling functions of $p_T$ the jet quenching
is mostly controlled by the induced FFs at $z$ very close to unity
\cite{BDMS_RAA}. 
For the baryon/meson ratios the parton energy loss is not
very important. It can give only a small effect due to the difference in 
the quark 
and gluon suppressions and FFs to baryons. We will not consider this effect.

\section{The collinear 
$|qg\rangle_{\{\bar{6}\}}$ and $|gg\rangle_{\{10\},\{\overline{10}\}}$
states and the anomalous baryon production}

The shift of the parton longitudinal momenta related to 
the induced gluon emission is not the only medium
effect in jet production in $AA$-collisions.
Due to the color exchanges between 
the fast moving partons and the QGP the parton system after escaping from 
the plasma may turn out to be in
the anomalous (forbidden in the vacuum) color states . 
We will consider only the two parton almost 
collinear $gq$ and $gg$ final states for quark and gluon
jets, respectively. 
In vacuum the $gq$ system is always in the $\{3\}$ 
color state, and the $gg$ system
in the $\{8_{f}\}$ color state.
However, after color exchanges with the QGP as shown in Fig.~1 
the  $gq$ system can be in the $\{3\}$, $\{\bar{6}\}$ and $\{15\}$
color states, and
the $gg$ system in the  $\{1\}$, $\{8_{f}\}$, $\{8_{d}\}$, $\{10\}$, 
$\{\overline{10}\}$, and
$\{27\}$ color states \footnote{In principle, 
for a non-local parton system one needs to fix a point where the color
representation is defined and introduce the Wilson lines connecting the
partons with this 
point (similarly to the formula (\ref{eq:10})).
However, in the perturbation theory (in a gauge with gluon
field vanishing at large distances, say,  in Feynman gauge) this complication
does not arise since the Wilson line factors can be replaced by unity.}.
These new color states, which are absent for jets in the vacuum,  
can violate the independence of the parton fragmentation, and the formulas
(\ref{eq:20}), (\ref{eq:30}) will not be valid anymore.
In principle, the collinear two parton configuration should violate the
independence of the parton fragmentation even in the vacuum (i.e., for hadron-hadron
collisions) in the string stage of hadronization. However, there such effects can effectively be included to the
phenomenological FFs at low virtuality scale. But this does not work for
$AA$-collisions when we have the new color states that are simply absent for the
vacuum fragmentation (if one neglects the higher-twist interactions between
comovers), and such states should be taken into account separately.

There is a set of new jet color chemistry phenomena related to 
production of the anomalous color states.
\begin{figure*}[htb]
\begin{center}
  \includegraphics[width=0.85\textwidth]{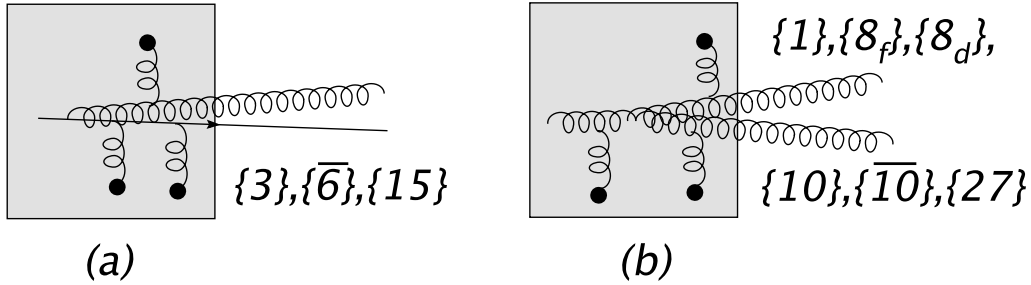}
\end{center}
\caption
{The $q\to gq$ (a) and $g\to gg$ (b) transitions in the medium and possible color states of
 the final two parton systems.}
\end{figure*}
The change of the total jet color charge may be important in the aspect of 
the large angle soft gluon emission which is sensitive namely to the 
total jet color charge. 
Evidently the appearance of the higher color states should increase the 
multiplicity in the fragmentation region as well.
This effect is similar to the modification of the parton branchings 
due to color exchanges
with the medium discussed recently in \cite{W_QM11} from the point of 
view of the large-$N_{c}$ limit. 
On the other hand, the appearance of the color singlet $gg$ state may lead
to jet events of the diffractive type with a rapidity gap between the bulk 
of soft hadrons and a fast moving meson system. The effect may be enhanced
for $\eta$ and $\eta^{'}$ mesons (or glueballs) due to the 
two-gluon Fock component in their 
wave functions \cite{Kroll_eta}. In principle, the experimental 
analysis of such events
would be of interest for the glueballs search. 
And what is interesting to us is that 
the production of the $\{\bar{6}\}$ $gq$ state and 
$\{10\}$, $\{\overline{10}\}$ $gg$ states can be a source of the CABF
since, 
as was noted in the Introduction, these two parton states 
can create the color strings decaying into hadron systems with a leading baryon
(or an anti-baryon for the $\{10\}$ $gg$ state).

\subsection{Basic concepts of the model}

Let us consider the CABF related to the collinear
$|qg\rangle_{\{\bar{6}\}}$ and $|gg\rangle_{\{10\},\{\overline{10}\}}$
configurations.
After escaping from the QGP there will be formed the sextet and decuplet color
flux tubes attached to these states.
The color
neutralization of the sextet color tube 
formed by the $|gq\rangle_{\{\bar{6}\}}$ system occurs through 
the tunnel creation from
the vacuum of two $q\bar{q}$ pairs (since the sextet color wave function
is a tensor $\Psi_{ij}$ with two color spinor indices).
It results in formation of the color
singlet primary heavy tube/prehadron state of unit baryon number. 
In Fig. 2a such a state is shown in terms of the 
elementary strings with triplet color flux and SJ.
The hard gluon as in the Lund model \cite{LUND} is represented by a kink
in the color string.
Similarly, for the 
$|gg\rangle_{\{10\},\{\overline{10}\}}$ systems 
the color neutralization of the decuplet color flux tubes occurs via 
tunnel production of three
$q\bar{q}$ pairs (since the decuplet color wave function is a tensor
$\Psi_{ijk}$ with three color spinor indices) \footnote{In principle, for the decuplet states the color
 neutralization of the decuplet flux tube can go through the tunnel production
 of two $gg$ pairs. However, the tunnel gluon production should be strongly
 suppressed due to large effective gluon mass in the QCD vacuum, and it is
 usually neglected in the quark-gluon string models.}.  
For the $|gg\rangle_{\{\overline{10}\}}$ state it results in formation of a prehadron state with the baryon number $B=1$ 
shown in Fig. 2b, and for the $|gg\rangle_{\{10\}}$ state a prehadron with
$B=-1$.
Of course, for formation of the prehadron states shown in Fig.~2 the parent
two parton states should be collinear enough. We will discuss the
corresponding conditions below.
The hadronization of the primary prehadrons shown in Fig.~2 via 
sequential decays should result in production of baryons in the jet
fragmentation region.
\begin{figure*}[htb]
\begin{center}
  \includegraphics[width=0.85\textwidth]{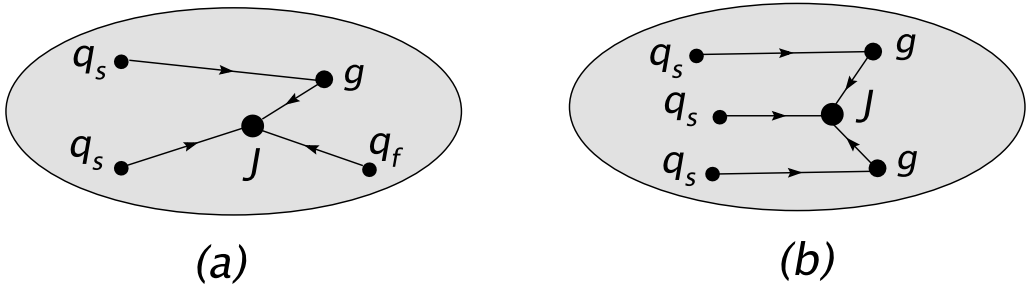}
\end{center}
\caption
{The primary heavy color flux tubes created by the 
$|qg\rangle_{\{\bar{6}\}}$ (a) and $|gg\rangle_{\{\overline{10}\}}$ (b)
  systems after color neutralization via the $q\bar{q}$ pairs production,
$q_{s}$ and $q_{f}$ are the low energy (created through the Schwinger tunnel
  mechanism) and high energy quarks, respectively. $J$ denotes the SJ.
}
\end{figure*}

An accurate description of the fragmentation of the color flux tubes formed by
the $|qg\rangle_{\{\bar{6}\}}$ and $|gg\rangle_{\{10\},\{\overline{10}\}}$
configurations is an extremely difficult problem.
We assume that the hadronization of the primary prehadrons shown in Fig.~2 can
be described as independent hadro\-nization of the color string configurations
shown in Fig.~2. Similarly to the Lund model \cite{LUND}, the hard gluon kinks in the primary prehadron states 
in Fig. 2 are assumed to split into the collinear $q\bar{q}$ pairs
with the pQCD $z$-distribution $W(z)\propto P_{qG}(z)\propto
[z^{2}+(1-z)^{2}]$ (here $z$ is the quark fractional momentum). 
In numerical calculations we 
neglect the gluon splitting into strange quarks.
After the $g\to \bar{q}q$ splitting
the primary prehadron states shown in Fig. 2 take the forms shown in Fig. 3.
\begin{figure*}[htb]
\begin{center}
  \includegraphics[width=0.85\textwidth]{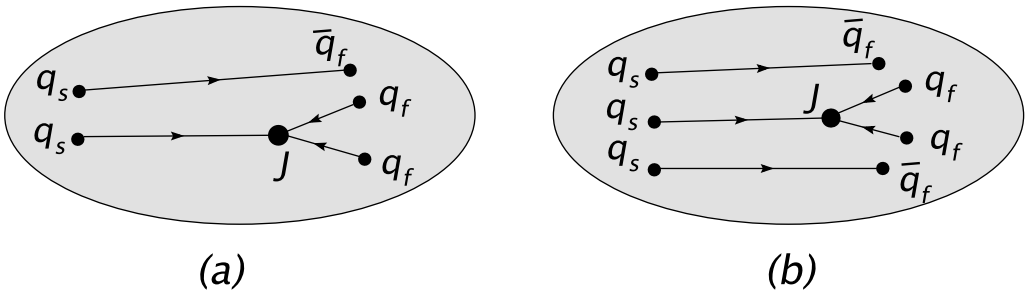}
\end{center}
\caption
{The same as in Fig.~2 after gluon splitting into $q\bar{q}$.
}
\end{figure*}

In the approximation of independent hadronization of the color strings
we can express the CABF corresponding to the configurations in Fig.~3 in terms of the diquark fragmentation
function $D_{B/D}$ for the color anti-triplet diquark $D_{\{\bar{3}\}}$
(for clarity we will omit the color index $\{\bar{3}\}$).
It
allows one to use as much as possible the information on the diquark fragmentation extracted from fitting the
baryon spectra in hadron collisions within the quark-gluon string model
\cite{MQGS1,MQGS2}. 
The independence of the hadronization
processes for the elementary color strings is the basic idea of the 
quark-gluon string
model \cite{MQGS1,MQGS2}. It is based on the topological 
expansion \cite{top-exp} scheme (for processes without baryons
the leading terms of the topological expansion are equivalent to the leading order
terms in the expansion in $1/N$ ($N=N_{c}$, $N_{c}/N_{f}=$const
\cite{V_QCD}). For the processes with baryons
which include the string configurations with the SJ the topological expansion
has not been rigorously established. But the idea that the triplet color 
strings
should hadronize independently for the configurations with the SJ as well 
is credible, and we use it as a reasonable working hypothesis for qualitative 
estimates of the CABF. We emphasize that the physical picture shown in 
Figs.~2,~3 in terms of the elementary
triplet color strings is not crucial for the fact that the $|qg\rangle_{\{\bar{6}\}}$ and
$|gg\rangle_{\{\overline{10}\}}$ states should produce the leading baryons.
Indeed, the recent lattice simulation of the color flux tubes performed in
\cite{lattice-tubes} demonstrates that the string tension 
for the triplet, sextet and decuplet color flux tubes is roughly proportional to
the number of the spinor color indices for the corresponding color
representations. For this reason one can expect that without reference to
the strings with the triplet color flow the
$q\bar{q}$ pair creation rates (which controls the resulting baryon
$z$-distribution) for the sextet and decuplet color flux tubes
should be approximately the same as for independent hadronization of the
elementary triplet strings.

Analytically in our approximation the anomalous baryon FF $D_{B/i}^{m}$,
which should be used in (\ref{eq:20}),
can be written as
\beq
D_{B/i}^{m}(z,Q,L)\approx \int_{z}^{1} \frac{dz'}{z'}D_{B/D}(z/z')
D_{D/i}(z',Q,L)\,
\label{eq:40}
\eeq
with $D_{D/i}(z,Q,L)$ being the probability distribution to find in the color
anomalous two
parton system
created by the parent parton $i$ with the virtuality $Q$ the diquark $D$ 
with fractional momentum $z$
(immediately after escaping from the QGP). The formula (\ref{eq:40}) is the
analogue to (\ref{eq:30}) written for a single parton under the 
approximation of independent parton fragmentation. 
Note that for our $|qg\rangle_{\{\bar{6}\}}$ and $|gg\rangle_{\{\overline{10}\}}$
configurations the diquark is automatically in the anti-triplet color state.
In our numerical computations we use the parametrization of the diquark to
proton fragmentation function
of the form $D_{p/D}(z)=Az^{m}$ with $m=5/2$ obtained in fitting the proton
spectra in hadron reactions within the quark-gluon string model 
in \cite{Kaid-Pisk}.

In calculating the $D_{D/i}$ we treat the diquark as being consisting  of 
the two fastest quarks in the prehadrons shown in Fig.~3. It is a natural
assumption since for the CABF we are interested 
in the hard region with $z$ close to
unity. Then we can calculate the diquark distribution treating perturbatively
the creation of the 
$|qg\rangle_{\{\bar{6}\}}$ and $|gg\rangle_{\{\overline{10}\}}$
configurations.
We calculate the diquark distribution 
describing the diquark as a two quark configuration with  
some invariant mass cutoff $M_{D}$. We will perform calculations for $M_{D}=1$
and $M_{D}=1.5$ GeV. These values seem to be reasonable in the light of the
calculations of the diquark masses within the Dyson-Schwinger equation which
give the diquark masses $\sim 0.8-1$ GeV \cite{diquark1,diquark2}. Since we
work with non-interacting quarks our parameter $M_D$ may be somewhat bigger
than the real diquark mass. For this reason the value $M_D=1.5$ does not look unrealistic.
Note that the mass cutoff at $M_D\sim m_N$ ensures that the baryon production
is not accompanied by 
production of the leading mesons with 
the fractional momentum $z\sim 1$. These mesons should inevitably shift 
the baryon distribution to the low-$z$ region which is not important from the point 
of view of the high-$p_{T}$ baryon yield. We will denote the 
diquark FF with this mass cutoff $D_{D/i}(z,M_{D})$ (hereafter we suppress
the arguments $Q$ and $L$). 

We emphasize that although the configurations with $M_D\gg m_N$ 
are not important for the baryon/meson ratios they, of course,
also lead to the CABF. Such configurations should enhance the baryon FFs 
in $AA$-collisions  as compared to the ordinary baryon FFs at $z\ll 1$.
The states with $M_D\gg m_N$ are produced when the radiated gluons
in Fig. 1 are very soft.
For the induced gluon radiation, which dominates the CABF for RHIC and LHC
conditions, most of the gluons have energies $\lsim 5$ GeV. 
After the $g\to \bar{q}q$ splitting the typical energy of the softest 
quarks in the $q_{f}q_{f}$ systems in the primary prehadron states 
in Fig. 3 will lie in the same energy range. 
For the very asymmetric $q_{f}q_{f}$ configurations
the SJ should have a rapidity close to that for the 
softest quark in the $q_{f}q_{f}$ pair (since it minimizes the string energy).
It will result in 
production of a baryon with energy about
the typical energy of the emitted gluons in Fig. 1, i.e., 
in the soft region $z\ll 1$.
The hadronization of the strings attached to the fastest quarks 
in the asymmetric $q_{f}q_{f}$ pairs in Fig. 3, 
which controls the FFs
in the hard region $z\sim 1$, should be similar to that for the ordinary 
vacuum fragmentation. 
In the present paper we concentrate namely on the role of the CABF in 
the baryon/meson ratios and will not consider the effect of the CABF 
on the FFs in the soft region.

\subsection{Calculation of the diquark distribution}

The main ingredient in our model for the CABF is the diquark
distribution $D_{D/i}$.
We illustrate its calculation for a quark jet.
For pure two parton $gq$ state , i.e. for 
$q\to |qg\rangle_{\{\bar{6}\}}$ transition without multiple gluon emission
we write $D_{D/q}$ as
\beq
D_{D/q}(z,M_{D})=\int_{0}^{1} dx dy \frac{dN_{gq}^{\{\bar{6}\}}}{dx}
W(y)\delta(z-z_{D})
\theta(M_{D}-M_{q_{1}q_{2}})\,.
\label{eq:60}
\eeq
Here $dN_{gq}^{\{\bar{6}\}}/dx$ is the gluon distribution for the 
$q\to |gq\rangle_{\{\bar{6}\}}$ transition,
with $x=p_{g}/p_{q}$ being the fractional gluon momentum, 
$y$ is the quark fractional 
momentum for $g\to \bar{q}q$ splitting, 
$z_{D}=1-x +xy$ is the total fractional
momentum of the fast quarks (we denote the quark which radiates the gluon
$q_{1}$ and the quark from the $g\to \bar{q}q$ splitting $q_{2}$),
and $M_{q_{1}q_{2}}$ is the invariant mass for the $q_{1}q_{2}$ system.

For the sake of simplicity we account for the transverse 
(to the jet axis) motion effects on the average, so to speak, 
via the effective transverse quark mass $m_{T}$
and approximate $M_{q_{1}q_{2}}$ by a one-dimensional formula 
\beq
M_{q_{1}q_{2}}^{2}=\frac{m_{T}^{2}}{z_{1}}+\frac{m_{T}^{2}}{z_{2}}\,,
\label{eq:70}
\eeq
where $z_{1}=(1-x)/z_{D}$, $z_{2}=xy/z_{D}$ are the fractional momenta 
of the quarks in terms of the $x$ and $y$ variables. Our choice of $m_{T}$ will be given below in Sec.~4.

The distribution function $dN_{gq}^{\{\bar{6}\}}/dx$ in (\ref{eq:60}) 
corresponds to the final 
$|gq\rangle_{\{\bar{6}\}}$ state escaping from the QGP.
The gluon in this state can be originally produced
either in the DGLAP stage due to the vacuum $q\to gq$ transition
without interaction with the medium 
or in the bulk QGP via the induced gluon emission. 
The distribution $dN_{gq}^{\{\bar{6}\}}/dx$ should be calculated including
both these mechanisms. We will refer to them as the DGLAP and induced 
mechanisms. We emphasize that in our case both of them are the medium
effects. Even for the DGLAP mechanism when the initial $gq$ system is 
produced in the triplet color state
the final $|gq\rangle_{\{\bar{6}\}}$ state escaping from the QGP 
emerges only due the subsequent color exchanges between the $gq$ system and 
the medium.

Let us now consider the incorporation of the multiple gluon emission,
which is neglected in the formula (\ref{eq:60}). The radiation of additional
gluons can occur in the  DGLAP stage (as was already noticed mostly at the
longitudinal scale $\sim 1$ fm) and in the subsequent induced gluon radiation
stage in the whole volume of the QGP.
An accurate description of the multiple gluon emission is a complicated
problem. However, it can be made in a reasonable approximation, which
is sufficient for our purposes here.
To incorporate the multiple gluon emission we use the 
fact that the CABF is dominated by a region of moderate $x$ around $x\sim 0.5$ 
(since only for such
configuration a hard and not very heavy diquark can be produced).
On the other hand, both the DGLAP and especially induced multiple
gluon radiation are mostly soft with $x\ll 1$. 
Thus the collinear two parton system determining the CABF and the soft 
gluons act, so to speak, in the different regions of the phase 
space.
For soft gluons
the two parton collinear $qg$ state with hard gluon should approximately act 
as a single fast quark. 
For this reason the effect of the Sudakov formfactor (related to
the virtual gluons) and real gluon
emission for the $gq$ state 
should roughly be similar to that for a single quark.

For a single quark the DGLAP and induced radiation lead simply to
replacement of the hard parton cross section by
\beq
\frac{d\sigma_{eff}(N+N\rightarrow j+X)}{d\pb_{T}^{j} dy}=
\sum_{i}\int_{0}^{1} \frac{dz}{z^{2}}
 D_{j/i}^{ind}(z,p_{T}^{i},L)
\times\frac{d\sigma_{DGLAP}(N+N\rightarrow i+X)}{d\pb_{T}^{i} dy}\,,
\hspace*{.5cm}
\label{eq:81}
\eeq
where
\beq
\frac{d\sigma_{DGLAP}(N+N\rightarrow j+X)}{d\pb_{T}^{j} dy} =
\sum_{i}\int_{0}^{1} \frac{dz}{z^{2}}
D_{j/i}^{DGLAP}(z,Q)
\frac{d\sigma(N+N\rightarrow i+X)}{d\pb_{T}^{i} dy}\,.
\label{eq:82}
\eeq
Here, as in (\ref{eq:31}), 
$ D_{j/i}^{ind}$ and $D_{j/i}^{DGLAP}$ denote the induced and DGLAP
partonic FFs,
and $p_{T}^{i}=p_{T}^{j}/z$.
This representation is written taking into account the time ordering
of the DGLAP and the induced radiation stages. 
Thus we can work with the formula (\ref{eq:60})
for pure two parton $gq$ state but at the same time we should replace 
the LO pQCD cross section by that given by Eq. (\ref{eq:81}).
In fact, for the baryon/meson ratio one can go one step 
further in simplification of the calculations.
It is based on the fact that the ratio $\sigma_{eff}/\sigma_{DGLAP}$
is close to the nuclear modification factor $R_{AA}$ for the meson cross
section. For this reason it is enough to replace the LO pQCD hard cross
section by the $\sigma_{DGLAP}$ (\ref{eq:82}), and at the same time to 
use for the denominator in the baryon/meson ratio the meson cross section for 
$NN$-collisions. 

We are fully aware that the above procedure does not treat
accurately the real and virtual multiple gluon emission in the hard region
itself. However, for the vacuum and induced $q\to gq$ distributions the
integral probability to find the gluon in the hard region (say, $x\gsim 0.2$ which
dominates the CABF) turns out to be smaller than unity. This says that the 
unitarity
effects related to the real and virtual multiple gluon emission are relatively
small, and the difference between these effects for a single quark and
the $gq$ state cannot give large errors.
In any case, presently, even for a single quark, there is no an accurate
method for evaluation of the multiple induced gluon emission. 

\subsubsection{Calculation of $dN_{gq}^{\{\bar{6}\}}/dx$}  

An accurate computation of 
$dN_{gq}^{\{\bar{6}\}}/dx$
even at one gluon level is a difficult problem. As already noticed above,
it should take into account the vacuum and induced $q\to gq$ 
transitions. It is important that for both these mechanisms the 
original $gq$ state is subjected to rotation in the color space due 
to multiple gluon exchanges with the medium.
Qualitative estimates show that for RHIC and LHC conditions the two parton 
system almost surely changes its color state in the QGP for 
jet energies $E\lsim 30-50$ GeV. For this
reason in this energy range 
one can expect that multiple
rescatterings should result in almost complete color randomization of the 
two parton system. 
Then the probability of the $\{\bar{6}\}$ color state is simply given by the
statistical weight factor 6/24. 

Let us first consider the distribution $dN_{gq}^{\{\bar{6}\}}/dx$ for 
the vacuum (DGLAP) creation of the $gq$ system, i.e.,
without interaction with the medium. 
We write the leading order $(x,\qb)$-distribution  (hereafter $\qb$ denotes
the momentum transverse to the jet axis) for 
the vacuum $q\to gq$ transition
in the form
\beq
\frac{dN_{vac}}{dxd\qb}=\frac{C_{F}\alpha_{s}}{2x\pi^{2}}[
1+(1-x)^{2}]\frac{q^{2}}{(q^{2}+\epsilon^{2})^{2}}\,.
\label{eq:80}
\eeq
Here $C_F$ is the quark color Casimir factor, $\epsilon^{2}=m_{q}^{2}x^{2}+m_{g}^{2}(1-x)$, 
$m_{q,g}$ are the quark and gluon masses. 
In our numerical computations 
we take  $m_{g}=0.75$ GeV. This value was obtained  
from the analysis of the low-$x$ proton structure function $F_{2}$ within the dipole BFKL equation
\cite{NZ_HERA}. It agrees well with the natural infrared cutoff 
for gluon emission $m_{g}\sim 1/R_{c}$, where 
$R_{c}\approx 0.27$ fm is the gluon correlation radius in the QCD vacuum
\cite{shuryak1}. 
For the quark mass we take $m_{q}=0.3$ GeV.
However, the value of the quark mass is not very important. 

Of course, the final (after escaping from the QGP) $(x,\qb)$-distribution 
for the $gq$ system created in the DGLAP stage, 
differs from the primordial one (\ref{eq:80}) due to the 
transverse momentum broadening in the QGP. 
We write it (we denote it by a subscript $br$) as a convolution
\beq
\frac{dN_{vac}^{br}}{dxd\qb}=
\int d\kb \frac{dN_{vac}}{dxd\kb}I_{br}(\qb-\kb)\,,
\label{eq:90}
\eeq 
where $I_{br}(\qb)$ is the distribution for the broadening in the internal
transverse momentum of the $gq$ system. The internal transverse momentum in
terms of the gluon and quark momenta reads $\qb=(1-x)\qb_{g}-x\qb_q$. 
We treat the $\qb$-broadening as an independent random walk of the quark and
gluon in transverse momenta characterized by their transport coefficients
\cite{qhat,BSZ}. Then the $I(\qb)$ has a Gaussian form
\beq
I_{br}(\qb)=\frac{1}{\pi \langle\qb^{2}\rangle}
\exp\left(-\frac{\qb^{2}}{\langle\qb^{2}\rangle}\right)\,,
\label{eq:910}
\eeq
\beq
\langle\qb^{2}\rangle=\int_{\tau_{0}}^{L} dl
[(1-x)^{2}\hat{q}_{g}(l)+x^{2}\hat{q}_{q}(l)]\,,
\label{eq:91}
\eeq
where 
$\hat{q}_{g}(l)$ and $\hat{q}_{q}(l)$ are the local gluon and quark transport
coefficients related by $\hat{q}_{g}(l)=\hat{q}_{q}(l)C_{A}/C_{F}$.
In (\ref{eq:91}) we have taken for the lower limit of the $l$-integral the QGP
formation time $\tau_{0}$ since the typical gluon formation time for the 
vacuum $q\to gq$ transition is of the order of $\tau_{0}$, and the $gq$ 
system should undergo scattering on the thermal partons immediately after 
formation of the QGP. In any case, the corresponding
inaccuracy cannot be big since the resulting $\langle\qb^{2}\rangle$ has only 
weak logarithmic dependence on the limits of the $l$-integration.

We write the resulting vacuum contribution to 
$dN_{gq}^{\{\bar{6}\}}/dx$ as 
\beq
\left.\frac{dN_{gq}^{\{\bar{6}\}}}{dx}\right|_{vac}=A_{\{\bar{6}\}}\int\limits_{q<q_{max}}
d\qb\frac{dN_{vac}^{br}}{dxd\qb}\,,
\label{eq:100}
\eeq
where we introduced the
color weight factor $A_{\{\bar{6}\}}=6/24$ of the anti-sextet state.   
The restriction $q<q_{max}$ is imposed to cut off the
non-collinear configurations which cannot produce the prehadron state
with two fast quarks located in the same cavity in the terminology of 
the MIT bag-model. For the transverse size of the prehadron 
cavity it is reasonable
to use the proton radius $R_{p}\sim 1$ fm (this value is also consistent
with the transverse size of the sextet and decuplet color tubes about $0.8$ fm
obtained in the lattice simulation \cite{lattice-tubes}).
Then for the parton energy $\sim 10-20$ GeV 
one can obtain $q_{max}\lsim 0.3-0.5$ GeV (we assume that for the radiated gluon
$x\sim 0.5$ and the size of the QGP is about 5 fm). 
Note that this transverse momentum region is reasonable from the point of view 
of the diquark mass as well, since it leads to $M_{D}\lsim 1-2$ GeV. 
A larger  value of $q_{max}$ (which is possible from the geometrical viewpoint
at higher energies) would give too heavy diquark which should 
have a smaller probability for fragmentation into a leading baryon.   

Let us now discuss the induced contribution 
to $dN_{gq}^{\{\bar{6}\}}/dx$.
An accurate evaluation of the $(x,\qb)$-dis\-tri\-bu\-tion for the induced gluon
emission (necessary for calculation of the $x$-distribution) even at one 
gluon level is problematic. In principle, 
the $(x,\qb)$-spectrum can be computed within the light-cone path integral
(LCPI) formalism \cite{LCPI,LCPI_rev}.
However, it requires solving a complicated two dimensional Schr\"odinger 
equation with a
non-central potential. In the present qualitative analysis, to avoid 
this complications,
we use the LCPI formalism only for calculation of the transverse 
momentum integrated
spectrum $dN_{ind}/dx$, and parametrize the transverse momentum dependence in
a Gaussian form. We assume that the mean squared momentum can be approximated
as a sum
\beq
\langle \qb^{2}\rangle=\langle \qb^{2}\rangle_{fz}+\langle
\qb^{2}\rangle_{ms}\,,
\label{eq:101}
\eeq
where the first term corresponds to the 
primordial internal momentum of the $gq$ system in the gluon formation zone
and the second one comes from the momentum accumulated due to 
multiple scatterings. The formation zone term can be written via the mean
squared transverse separation between gluon and quark as  
$\langle \qb^{2}\rangle_{fz}\sim 4/\langle \rho_{gq}^{2}\rangle$. From the
Schr\"odinger diffusion relation one can obtain $\langle
\rho_{gq}^{2}\rangle\sim 2L_{typ}/x(1-x)E$. Here
$L_{typ}= \mbox{max}(L_{QGP},L_{f})$, where $L_{f}$ is the gluon formation
length.
For induced radiation the gluon formation length is approximately 
$L_{f}\sim 2x(1-x)E S_{LPM} /\epsilon_{pl}^{2}$ \cite{LCPI_rev}, 
where $S_{LPM}$ is the
Landau-Pomeranchuk-Migdal (LPM) suppression factor, the index $pl$ 
of $\epsilon_{pl}$ indicates
that it should be calculated with parton masses replaced by 
the corresponding quasiparticle masses in the plasma. We use the quasiparticle
masses $m_{q}=0.3$ and $m_{g}=0.4$ GeV obtained within the quasiparticle 
model of the QGP in \cite{LH}. The LPM suppression for RHIC and 
LHC conditions is not
very strong. Typically the LPM effect suppresses the gluon spectrum by a
factor $\sim 1.5-3$. For moderate $x$, important for the CABF, and parton energy
region $E\sim 5-30$ the $L_f$ turns out to be of the order of the plasma size
or even larger.
This says that, strictly speaking, one cannot separate the effects of the
primordial internal momentum in the gluon formation zone and that due to
multiple scatterings. However, for our rough estimates it is not very
important since numerically the effect of the primordial momentum is
relatively small, and the width of the transverse momentum distribution is
mostly controlled by multiple scatterings.

Note that in our treatment of the $\qb$-broadening (both for the vacuum and
induced $gq$ states) we neglected the contribution of 
the transverse momentum kicks related to the soft gluon emission. 
An accurate evaluation of this effect is complicated.
However, it should not change drastically our results. First of all, the
number of elastic kicks is considerably larger than that from the radiated
soft gluons. Also, the kicks from the soft gluon emission mostly occur 
before the
formation of the $gq$ system with a hard gluon (since the gluon formation
length is $\propto x$) and do not affect the $\qb$-distribution at all.
This is especially true for the induced mechanism, for which multiple
gluon emission is very soft. As will be seen from our numerical results
namely the induced mechanism dominates the CABF.

In connection with Eq. (\ref{eq:60}) 
it is perhaps worth noting that although  
for the dominating induced mechanism the $x$-distribution
$dN_{gq}^{\{\bar{6}\}}/dx$ is concentrated in a sharp peak
at small $x$ with a width $\propto 1/E$ it does not generate
a strong energy dependence of the CABF. The strong energy dependence 
could naively be expected from the decrease with the jet energy of 
the overlap of
the region with $M_{D}\lsim m_N$ with the low-$x$ peak in 
$dN_{gq}^{\{\bar{6}\}}/dx$. In reality, the low-$x$ peak in the gluon
distribution is practically not important for the baryon yield for $p_{T}
\gsim 5-10$ GeV. Our numerical calculations show that in the
region $x\gsim 0.2-0.3$, which dominates the CABF for $M_{D}\sim m_N$,
the distribution $dN_{gq}^{\{\bar{6}\}}/dx$  already has a quite 
smooth $x$-dependence (approximately $\propto
1/x^{\alpha}$ with $\alpha\sim 1-1.5$  for the jet energy $E\sim 5-50$ GeV). 
The same is true for the $dN_{gg}^{\{\overline{10}\}}/dx$
entering Eq. (\ref{eq:110}) for the gluon jets. 
As already noticed above, the low-$x$ peak in the gluon distribution
is, of course, important for the enhancement of the baryon FFs in the
the soft region $z\ll 1$. This effect, related to the very asymmetric 
$q_{f}q_{f}$ configuration in Fig. 3, is not described by Eqs. 
(\ref{eq:60}) and (\ref{eq:110}). But the baryon yield
at high $p_{T}$, which we are interested in, is practically insensitive to the 
baryon FFs at small $z$.

\subsubsection{The gluon jets}

The baryon production for gluon jet is treated in a similar way as a 
three-step
process that now involves first the production of the $gg$ pair in the
decuplet/anti-decuplet color state, followed by the diquark formation 
(after splitting of hard gluons 
into $q\bar{q}$ pairs and formation of the configuration shown in Fig.~3b), 
and final diquark fragmentation to the observed baryon.
The analog of Eq. (\ref{eq:60}) for the state shown in Fig.~3b has the form
\beq
D_{D/g}(z,M_{D})=\frac{1}{2!}\int_{0}^{1} dx dy_{1}dy_{2} 
\frac{dN_{gg}^{\{\overline{10}\}}}{dx} 
W(y_{1})W(y_{2})\delta(z-z_{D})
\theta(M_{D}-M_{q_{1}q_{2}})\,, \hspace*{.5cm}
\label{eq:110}
\eeq
where  now $z_{D}=(1-x)y_{1}+xy_{2}$, 
the factor $1/2!$ accounts for the gluon identity,
and the light-cone fractional momenta in
the formula (\ref{eq:70}) for the invariant mass $M_{q_{1}q_{2}}$ read 
$z_{1}=(1-x)y_{1}/z_{D}$ and $z_{2}=xy_{2}/z_{D}$.
Similarly to the quark case,
for the $(x,\qb)$-distribution for the vacuum $g\to gg$ transition  
we use the leading order formula
\beq
\frac{dN_{vac}}{dxd\qb}=\frac{C_{A}\alpha_{s}}{\pi^{2}}\Big[
\frac{1-x}{x}+\frac{x}{1-x}+x(1-x)\Big]\frac{q^{2}}{(q^{2}+\epsilon^{2})^{2}}\,,
\label{eq:120}
\eeq
where now $C_A$ is the gluon color Casimir factor, and $\epsilon^{2}=m_{g}^{2}(1-x+x^{2})$. 
The statistical weight factor for the $\{\overline{10}\}$ color state is now
$A_{\{10\}}=10/64$. 

\subsubsection{The energy dependence beyond the color randomization
  approximation}

We conclude this section with a general comment about  
the validity region of the color randomization 
approximation and possible energy dependence of the CABF
above the color randomization window. In the approximation of color 
randomization the diquark distribution $D_{D/i}$ has a weak 
energy dependence.
However, the CABF is in general a higher-twist effect and 
is expected to have disappeared in the limit of high parton energies.
This phenomenon is realized in the following way.
For a given maximal internal momentum $q_{max}$ the transverse separation
between partons in the two parton system vanishes 
$\propto 1/E$ at $E\to \infty$.
Since for the small size two parton states the probability of changing the
color state is proportional to the size squared, the
probability of production of the sextet $gq$ and decuplet/anti-decuplet $gg$
states should vanish $\propto 1/E^{2}$ in the high energy limit. This says
that the approximation of the color randomization should be violated at
sufficiently high energies. Our qualitative estimates show that 
for $E\lsim 20-30$ GeV the approximation of color randomization should still be
reasonable. And then in this region the higher-twist nature of the CABF 
should not manifest itself.
Of course, 
one cannot exclude that the window of the
color randomization may turn out to be narrower and the dynamical effects 
lead to some energy dependence of the weight of the $\{\bar{6}\}$ and
$\{10\}/\{\overline{10}\}$ states even at $E\sim 10-30$ GeV.
But they should not change drastically the magnitude of the CABF in
this energy region.  

In principle, the energy dependence of the color anomalous two
parton configurations can be evaluated in the
dynamical approach based on the evolution operator 
for the density matrix for the two parton states 
\cite{dens-matrix1,dens-matrix2}.
However, our aim here is only to understand the potential importance of the
CABF, and it is outside the scope of this paper to study 
accurately its energy dependence.

\section{Numerical results}

We performed the computations of the vacuum two parton distributions
(\ref{eq:80}), (\ref{eq:120}) with 
the running $\alpha_s$ frozen at the value $\alpha_{s}^{fr}=0.7$.
This value was previously obtained 
by fitting the data on nuclear shadowing \cite{NZ} and  proton structure function
$F_{2}$ at low $x$
within the dipole BFKL equation \cite{NZ_HERA}. 
A similar value of 
$\alpha_{s}^{fr}$ follows from the relation 
$
\int_{\mbox{\small 0}}^{\mbox{\small 2 GeV}}\!dQ\frac{\alpha_{s}(Q^{2})}{\pi}
\approx 0.36 \,\,
\mbox{GeV}\,
$
obtained in \cite{DKT} from the analysis of the heavy quark energy 
loss in vacuum.
We calculate the hard parton cross sections  using the LO 
pQCD formula. To simulate the higher order $K$-factor
we take for the virtuality scale in $\alpha_{s}$ the value 
$cQ$ with $c=0.265$ as in the PYTHIA event generator \cite{PYTHIA}.

The DGLAP parton FFs  entering (\ref{eq:82}) were 
computed with the
help of the PYTHIA event generator \cite{PYTHIA}.
The $x$-distribution for the induced gluon emission has been computed within the 
LCPI formalism \cite{LCPI} using 
the  method elaborated in \cite{Z04_RAA}. In computing the induced gluon
emission we use the running $\alpha_s$ frozen at the value
$\alpha_{s}^{fr}=0.6$. The suppression of the in-medium $\alpha_{s}^{fr}$, which may be due
to the thermal effects, is supported by the analysis of the RHIC and LHC data 
on the nuclear modification factor $R_{AA}$ \cite{raa08,raa11}.

We approximate the effective transverse quark mass entering (\ref{eq:70}) 
as $m_{T}=\sqrt{m_{q}^2+\langle p_{q}^{2}\rangle }$, where $\langle
p_{q}^{2}\rangle$ is the mean squared quark transverse momentum. 
We estimated $\langle p_{q}^{2}\rangle$ for the configurations
in which all the gluon momentum in the $g\to q\bar{q}$ splitting
is transferred to the quark (such configurations dominates the CABF). 
Then a reasonable estimate is   
$\langle p_{q}^{2}\rangle\approx q_{max}^{2}/2$.

We calculated the $p/\pi^{+}$ and $\bar{p}/\pi^{-}$ ratios
for the central $Au+Au$ collisions at $\sqrt{s}=200$
GeV at RHIC, and for the $Pb+Pb$ central collisions at $\sqrt{s}=2.76$ TeV  
at LHC. The $\qb$-broadening has been calculated using  
the Bjorken model \cite{Bjorken} of the QGP evolution
with $T_{0}^{3}\tau_{0}=T^{3}\tau$. We take $\tau_{0}=0.5$ fm, and 
$T_{0}=300$ MeV for RHIC, and $T_{0}=400$ MeV for LHC.
These initial temperatures were obtained using the entropy/multiplicity ratio
$dS/dy\Big{/}dN_{ch}/d\eta=7.67$ 
from \cite{BM-entropy}.
In calculating the $\qb$-broadening for the typical parton path length in the
QGP in Eq. (\ref{eq:91}) we take $L=5$ fm which is
a reasonable value for the central collisions \cite{Z04_RAA}.
For the multiple scattering term $\langle \qb^{2}\rangle_{ms}$
in the $\qb$-broadening formula (\ref{eq:101}) for the induced 
mechanism we take for the 
minimal length $L_{min}=2$ fm (instead of $\tau_{0}$ as in Eq. (\ref{eq:91}) for the vacuum
mechanism). It is done to account for the fact that for
the induced mechanism the gluon radiation is distributed in the region of the
size $\sim L_{f}$. However, it gives a relatively small suppression of the
$\qb$-broadening since the dependence of $\langle \qb^{2}\rangle_{ms}$ on
$L_{min}$ is only logarithmic. For the gluon transport coefficient we take 
$\hat{q}_{g}=0.3$ GeV$^{3}$ at $T=250$ MeV. This value is approximately
consistent with accurate treatment of the $p_T$-broadening for our
parametrization of $\alpha_s$.

\begin{figure} [ht]
\begin{center}
\epsfig{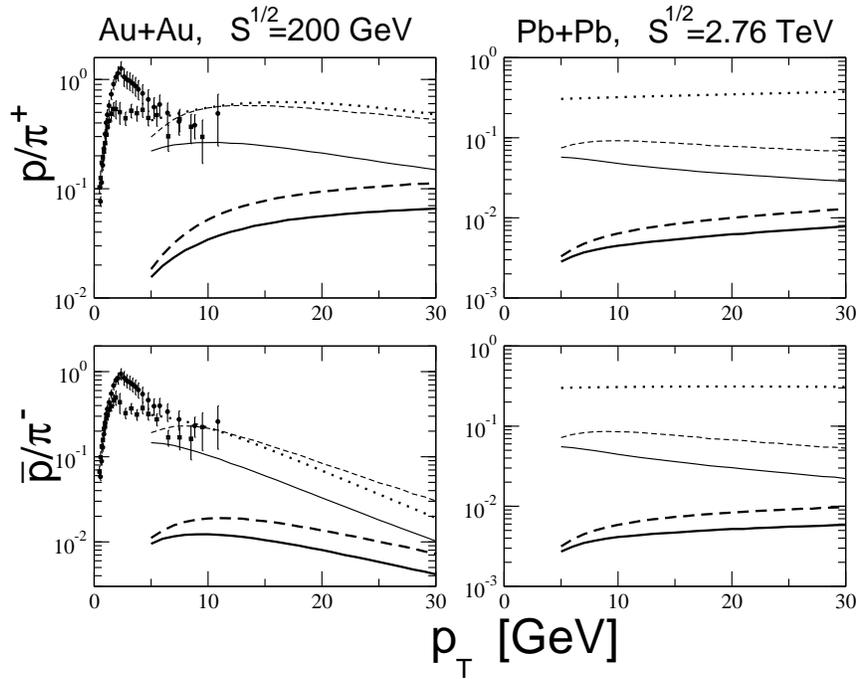}
\end{center}
\caption[.]
{
The $p/\pi^{+}$ and $\bar{p}/\pi^{-}$ ratios for 
$Au+Au$ at $\sqrt{s}=200$ GeV and $Pb+Pb$ at $\sqrt{s}=2.76$ TeV 
(central collisions). The thick lines show our results for
the CABF due to the $gq$ and $gg$ systems created in the 
DGLAP stage obtained for $M_{D}=1$ GeV (solid) and 
$M_{D}=1.5$ GeV (dashed), and
the thin lines the same for the CABF
due to the two parton systems created in the 
induced stage.
The dotted lines show the results for $NN$-collisions obtained with the AKK
FFs \cite{AKK}.
The experimental points in the left panels show the data from STAR for
$Au+Au$ 0-12\% central collisions \cite{STAR2} (circles) and $d+Au$ \cite{STAR_d} (squares) collisions.
}
\end{figure}

The results for $p/\pi^{+}$ and $\bar{p}/\pi^{-}$ ratios 
obtained for $q_{max}=0.3$ GeV, $m_{q}=0.3$ GeV, and
$M_{D}=1$ and $M_{D}=1.5$ GeV are plotted in Fig. 4. 
The results are shown separately for the vacuum mechanism and the induced one.
One can see that the induced mechanism dominates. This is mostly due to
the fact that from the beginning of the formation of the $gq$ and $gg$
systems their configurations are more collinear for the induced
mechanism.
The sensitivity to the diquark mass becomes stronger at larger $p_{T}$.
From Fig.~4 we see that for RHIC the effect of the CABF  is bigger than for LHC.
It is determined by two facts: one is the $p_{T}$-dependence of the parton
cross sections which is steeper at RHIC. The other is 
relative decrease of the quark yield at LHC which dominates the CABF at RHIC.
Note that the fact that $p/\pi^{+}$ and $\bar{p}/\pi^{-}$ ratios numerically are not small
does not mean that the probability to find a proton/anti-proton in parton fragmentation
is large. In fact, the baryon fraction is very small $\sim 1-5$\%. But for
baryons the dominating values of $z$ in (\ref{eq:20}) are bigger than for mesons, and due
to the steep fall-off of the parton cross section it enhances strongly the 
numerator in the baryon/meson ratio.     

The results shown in Fig.~4 have been obtained with the gluon splitting
function $W(z)=3[z^{2}+(1-z)^{2}]/2$ which in principle
has no theoretical justification for the gluon kinks in our model. To 
understand the sensitivity to the form of $W(z)$ we also
performed the calculations for the ansatz $W(z)=30z^2(1-z)^2$ when
the quark and anti-quark have almost exactly half the gluon momentum.
For this parametrization the $p/\pi^{+}$ and $\bar{p}/\pi^{-}$ ratios
became only $\sim 15-30$\% smaller. This demonstrates that the ansatz for
$W(z)$ does not play a crucial role.

To get insight into how strongly the CABF can modify the picture based 
on independent parton fragmentation with the ordinary vacuum FFs 
in Fig.~4 we also plotted the baryon/meson ratios obtained with the 
AKK baryon FFs \cite{AKK}. As one can see 
for RHIC the contribution of the CABF is of the same order as that for the
vacuum AKK FFs. For LHC the ratio CABF/AKK becomes somewhat smaller, but the
effect of the CABF is not negligible. Thus, our estimates demonstrate that the
CABF can potentially lead to a substantial violation of the scheme based on the 
ordinary FFs.
Note that comparing the CABF with the ordinary vacuum baryon fragmentation,
one has to keep in mind that the vacuum baryon FFs themselves have considerable
uncertainties. For example, the proton yield for the KKP FFs \cite{KKP} is
much smaller than that for the AKK FFs \cite{AKK}, and the relative effect of
the CABF is bigger in this case.

In Fig.~4 we also plotted the STAR experimental data for $Au+Au$ \cite{STAR2} and $d+Au$
\cite{STAR_d} collisions. Since the medium effects should be small for 
$d+Au$ collisions, the difference between the $Au+Au$ and $d+Au$ data
may be attributed to the medium effects that are of interest in our analysis. As one can see at $p_{T}\sim
5-10$ GeV our estimates for the contribution of the CABF are of the same order as  
the difference between the data for $Au+Au$ and $d+Au$ collisions.
However, the experimental errors are large and definite conclusion 
cannot be drawn from this comparison. 
At small transverse momenta $p_{T}\sim 2-3$ GeV
the difference between the data for $Au+Au$ and $d+Au$ is clearly seen, and it
is of the same order 
of magnitude as our estimates of the CABF for $p_{T}\sim 5$ GeV. 
But it can hardly be taken seriously since even at $p_{T}\sim 5$ GeV our approach 
is probably of limited applicability, and extrapolation of our results to
$p_{T}\sim 2-3$ GeV would be highly speculative.

The results presented in Fig.~4 for $m_q=0.3$ GeV can be viewed as
conservative estimates. For massless quarks 
the effect is enhanced by a factor of $\sim
1.5-3$. A similar enhancement occurs for $q_{max}=0.5$ GeV (which does not seem
unrealistic).
Thus, although our analysis is not expected to give accurate
quantitative predictions it clearly indicates that the CABF can be potentially
important at RHIC and LHC energies. And this mechanism deserves more careful
investigation. In particular, it is highly desirable to perform an 
analysis of the CABF beyond the color randomization approximation for better
understanding its energy dependence.
We postpone it for further work.

\section{Conclusions}

Besides the well known suppression of high-$p_{T}$ hadron yields in
$AA$-collisions, the color exchanges between jets and the QGP can lead 
to several other new phenomena. 
One of these jet color chemistry effects
is the anomalous contribution to the baryon yield \cite{M98}. 
We have estimated the anomalous baryon production due to this mechanism
accounting for formation of the collinear two parton 
$|gq\rangle_{\{\bar{6}\}}$ (for quark jets)
and $|gg\rangle_{\{\overline{10}\},\{10\}}$ states (for gluon jets). The
hadronization of these states can result in production of leading baryons.
Conceptually this mechanism is similar to the gluon mechanism of the 
baryon number flow in the $B\bar{B}$-annihilation and baryon
stopping  via gluon exchanges producing in a baryon/anti-baryon the  
two quark $qq/\bar{q}\bar{q}$-configurations in 
the $\{6\}/\{\bar{6}\}$ color states 
and the three quark $qqq/\bar{q}\bar{q}\bar{q}$-configurations in the 
$\{10\}/\{\overline{10}\}$ color states 
\cite{DEC,Y1988,Y1989,bflow}.

We have estimated the contribution of our mechanism to 
the $p/\pi^{+}$ and $\bar{p}/\pi^{-}$
ratios for the two parton anomalous states produced via 
the DGLAP and induced gluon emission. We found that the induced mechanism
dominates. The effect is stronger for quark jets.
For RHIC our estimates give the contribution 
to the $p/\pi^{+}$ and $\bar{p}/\pi^{-}$
ratios of the same
order as the ordinary vacuum fragmentation for AKK \cite{AKK} FFs.
At $p_{T}\sim 5$ the anomalous contributions to the  
$p/\pi^{+}$ and $\bar{p}/\pi^{-}$
ratios are comparable with the medium modification of these ratios 
observed by the
STAR Collaboration \cite{STAR2}.
For LHC the anomalous contribution becomes smaller than the vacuum one, but
not negligible. 

Of course, our estimates are extremely crude. But they indicate 
that the color anomalous baryon fragmentation can be potentially
important at RHIC and LHC energies. 
Contrary to the SJ loops \cite{SJL1} and recombination
\cite{REC_H,REC_F,REC_L,REC_rev} models
of the baryon anomaly the mechanism discussed in the present paper 
is purely of fragmentation type, and can be important at higher $p_{T}$.

\vspace {.7 cm}
\noindent
{\large\bf Acknowledgements}

\noindent
The work of BGZ  is supported 
in part by the 
Laboratoire International Associ\'e "Physique Th\'eorique et Mati\`ere Condens\'ee" (ENS-Landau)
and the grant SS-6501.2010.2.

\vskip .5 true cm


\begin{thebibliography}{99}

\bibitem{MT}
M.J.~Tannenbaum,
Rept.~Prog.~Phys. {\bf 69}, 2005 (2006), and references therein.

\bibitem{PHENIX1} 
S.S. Adler, {\em et al.} [PHENIX Collaboration], 
Phys.\ Rev.\ Lett. {\bf 91}, 172301 (2003).
 
\bibitem{STAR1} 
J. Adams, {\em et al.} [STAR Collaboration], 
Phys.\ Rev.\ Lett. {\bf 92}, 052302 (2004).

\bibitem{STAR2}
B.I. Abelev {\em et al.}  [STAR Collaboration],
Phys.~Rev.~Lett. {\bf 97}, 152301 (2006).


\bibitem{SJL1}
I. Vitev and M. Gyulassy,
Phys. Rev. C{\bf 65}, 041902 (2002).

\bibitem{SJL2}
V.~T.~Pop, M.~Gyulassy, J.~Barrette, C.~Gale, X.~N.~Wang, and N.~Xu,
Phys.\ Rev.\  C{\bf 70}, 064906 (2004).


\bibitem{RV1}
G.C. Rossi and G. Veneziano,
Nucl. Phys. B{\bf 123}, 507 (1977).

\bibitem{RV2}
G.C. Rossi and G. Veneziano,
Phys. Rep.  {\bf 63}, 149 (1980).

\bibitem{top-exp}
G.~Veneziano, 
Phys.~Lett. B{\bf 52}, 220 (1974); Nucl.~Phys. 
B{\bf 74}, 365 (1974).


\bibitem{Y-shape1}
T.T.~Takahashi, {\em et al.}, 
Phys.~Rev.~Lett. {\bf 86}, 18 (2001);
T.T.~Takahashi, {\em et al.},
Phys.~Rev. D{\bf 65}, 114509 (2002).

\bibitem{Y-shape2}
F.~Bissey, {\em et al.},
Phys.~Rev. D{\bf 76}, 114512 (2007).

\bibitem{REC_H} 
R.C. Hwa and C.B. Yang, 
Phys.\ Rev.\ C{\bf 67}, 064902 (2003).  

\bibitem{REC_F} 
R.J. Fries, B. Muller, C. Nonaka, and S.A. Bass, 
Phys.\ Rev.\ Lett. {\bf 90}, 202303 (2003).

\bibitem{REC_L} 
V. Greco, C.M. Ko, and P. L\'evai, 
{Phys.\ Rev.\ Lett.} {\bf 90}, 202302 (2003).


\bibitem{REC_rev}
R.J. Fries, V. Greco, and P. Sorensen,
Ann. Rev. Nucl. Part. Sci. {\bf 58}, 177 (2008) [arXiv:0807.4939 [nucl-th]].

\bibitem{M98} B.G.~Zakharov, 
Proceedings of the 33rd Rencontres de Moriond:
QCD and High Energy Hadronic Interactions,
Les Arcs, France, March 21-28, 1998, pp. 465-469
[arXiv:hep-ph/9807396].


\bibitem{BSZ} 
         R.~Baier, D.~Schiff, and B.~G.~Zakharov,
         Annu.\ Rev.\ Nucl.\ Part.\ Sci.\ {\bf 50}, 37  (2000).

\bibitem{DEC}
B.Z.~Kopeliovich and B.G.~Zakharov,
Phys.~Lett. B{\bf 211}, 221 (1988).


\bibitem{Y1988}
B.G. Zakharov and B.Z. Kopeliovich,
Sov.~J.~Nucl.~Phys. {\bf 48}, 136 (1988).


\bibitem{Y1989}
B.G. Zakharov and B.Z. Kopeliovich,
Sov.~J.~Nucl.~Phys. {\bf 49}, 674 (1989).

\bibitem{bflow}
B.Z. Kopeliovich and B.G. Zakharov,
Z.~Phys. C{\bf 43}, 241 (1989).

\bibitem{ann_rev}
B.G. Zakharov and B.Z. Kopeliovich,
Sov.~J.~Part. Nucl. {\bf 22}, 67 (1991).

\bibitem{ISR}
B.~Aper {\em et al.}, 
Nucl.~Phys. B{\bf 100}, 237 (1975);
T.~Akesson {\em et al.}, 
Nucl.~Phys. B{\bf 228}, 409 (1983).

\bibitem{MQGS1}
A.B.~Kaidalov,
JETP Lett. {\bf 32}, 474 (1980);
Phys.~Lett. B{\bf 116}, 459 (1982).

\bibitem{MQGS2}
A.~Capella and J.~Tran Thanh Van,
Phys.~Lett. B{\bf 114}, 450 (1982).

\bibitem{LUND}
B. Andersson {\em et al.},
Phys.~ Rep. {\bf 97}, 31 (1983).


\bibitem{GN}
E.~Gotsman and S.~Nussinov, 
Phys.~Rev. D{\bf 22}, 624 (1980).

\bibitem{BDMS_RAA}
R.~Baier, Yu.L.~Dokshitzer, A.H.~Mueller, and
D.~Schiff, JHEP {\bf 0109}, 033 (2001). 


\bibitem{raa08}
B.G.~Zakharov, JETP Lett. {\bf 88}, 781 (2008)
[arXiv:0811.0445 [hep-ph]].

\bibitem{Nussinov}
A.~Gasher, H.~Neuberger, and S.~Nussinov, 
Phys.~Rev. D{\bf 20}, 179 (1979).

\bibitem{W_QM11}
A.~Beraudo, J.G.~Milhano, and U.A.~Wiedemann,
arXiv:1107.1080 [hep-ph]; 
arXiv:1109.5025 [hep-ph].

\bibitem{Kroll_eta}
P.~Kroll and K.~Passek-Kumericki,
Phys. Rev. D{\bf 67}, 054017 (2003).

\bibitem{V_QCD}
G.~Veneziano, 
Nucl.~Phys. 
B{\bf 117}, 519 (1976).

\bibitem{lattice-tubes}
N.~Cardoso, M.~Cardoso, and P.~Bicudo, arXiv:1108.1542[hep-lat]. 



\bibitem{Kaid-Pisk}
A.B.~Kaidalov and O.I.~Piskunova,
Z.~Phys. C{\bf 30}, 145 (1986);
G.H.~Arakelian , A.~Capella , A.B.~Kaidalov, and Yu.M. Shabelski,
Eur.~Phys.~J. C{\bf 26}, 81 (2002)
[arXiv:hep-ph/0103337].

\bibitem{diquark1}
P.~Maris,
Few Body Syst. {\bf 32}, 41 (2002).


\bibitem{diquark2}
C.D. Roberts,
Prog.~Part.~Nucl.~Phys. {\bf 61}, 50 (2008), and references therein.



\bibitem{NZ_HERA}
N.N.~Nikolaev and B.G.~Zakharov,
Phys. Lett. B{\bf 327}, 149 (1994). 


\bibitem{shuryak1}
E.V.~Shuryak, Rev.\ Mod.\ Phys. {\bf 65}, 1 (1993).



\bibitem{qhat}
R.~Baier, Y.L.~Dokshitzer, A.H.~Mueller, S.~Peigne, and D.~Schiff,
Nucl. Phys. B{\bf 484}, 265 (1997).




\bibitem{LCPI}
B.G. Zakharov, JETP\ Lett. {\bf 63}, 952 (1996); {\em ibid}
{\bf 65}, 615 (1997);
{\bf 70}, 176 (1999).

\bibitem{LCPI_rev}
B.G. Zakharov, 
Phys.\ Atom.\ Nucl. {\bf 61}, 838 (1998).

\bibitem{LH}
P.~L\'evai and U.~Heinz,
Phys.\ Rev.\ C{\bf 57}, 1879 (1998).


\bibitem{dens-matrix1}
N.N.~Nikolaev, W.~Schafer, and B.G.~Zakharov,
Phys.~Rev. D{\bf 72}, 114018 (2005) [arXiv:hep-ph/0508310].


\bibitem{dens-matrix2}
N.N.~Nikolaev, W.~Schafer, B.G.~Zakharov, and V.R.~Zoller, 
Phys.~Rev. D{\bf 72}, 034033 (2005) [arXiv:hep-ph/0504057].

\bibitem{NZ}
N.N.~Nikolaev and B.G.~Zakharov,
Z.~Phys. C{\bf 49}, 607 (1991). 

\bibitem{DKT}
Yu.L.~Dokshitzer, V.A.~Khoze, and S.I.~Troyan,
Phys.\ Rev. D{\bf 53}, 89 (1996).


\bibitem{PYTHIA}
T.~Sjostrand, L.~Lonnblad, S.~Mrenna, and  P.~Skands,
arXiv:hep-ph/0308153.


\bibitem{Z04_RAA}
B.G.~Zakharov, JETP Lett. {\bf 80}, 617 (2004) [arXiv:hep-ph/0410321].

\bibitem{raa11}
B.G.~Zakharov, JETP Lett. {\bf 93}, 683 (2011) [arXiv:1105.2028 [hep-ph]].

\bibitem{Bjorken}
J.D.~Bjorken, 
Phys.\ Rev. D{\bf 27}, 140 (1983).

\bibitem{BM-entropy}
B.~M\"uller and K.~Rajagopal,
Eur. Phys. J. C{\bf 43}, 15 (2005).

\bibitem{AKK}
S.~ Albino, B.A.~Kniehl, and G.~Kramer,
Nucl. Phys. B{\bf 803}, 42 (2008).


\bibitem{KKP}
B.A.~Kniehl, G.~Kramer, and B.~Potter, 
Nucl.\ Phys. B{\bf 582}, 514 (2000).


\bibitem{STAR_d}
J.~Adams, {\em et al.} [STAR Collaboration],
Phys. Lett. B{\bf 637}, 161 (2006).




\end{thebibliography}
\end{document}